\documentclass[12pt]{iopart}

\usepackage{amssymb}
\usepackage{graphicx}
\usepackage{tabularx}
\usepackage{color}
\usepackage{epsfig}
\usepackage{psfrag}
\usepackage{epstopdf}
\usepackage{iopams}
\usepackage{indentfirst}
 \usepackage{pdfpages}

\begin{document}

\title{Effects of breaking vibrational energy equipartition on measurements of temperature in macroscopic oscillators subject to heat flux} 

\author{Livia Conti$^{1,*}$, Paolo {De Gregorio}$^{1,2,\dag}$, Gagik Karapetyan$^1$, Claudia Lazzaro$^1$, Matteo Pegoraro$^1$, Michele Bonaldi$^{3,4}$ and Lamberto Rondoni$^{2,5}$}
\address{$^1$ Istituto Nazionale di Fisica Nucleare, Via Marzolo 8, I-35131 Padova, Italy.}
\address{$^2$ Dip. Scienze Matematiche, Politecnico di Torino, Cso Duca degli Abruzzi 24, 10129 Torino Italy.}
\address{$^3 $ Institute of Materials for Electronics and Magnetism, Nanoscience-Trento-FBK Division,  38123 Povo (Trento), Italy.}
\address{$^4$ Istituto Nazionale di Fisica Nucleare, Gruppo Collegato di Trento, Sezione di Padova, 38123 Povo (Trento), Italy.}
\address{$^5$ Istituto Nazionale di Fisica Nucleare, Sezione di Torino, Via P. Giura 1, 10125 Torino, Italy.}
\ead{$^*$ livia.conti@pd.infn.it}
\ead{$^\dag$ paolo.degregorio@pd.infn.it}

\date{Sept 04, 2013}

\begin{abstract}

When the energy content of a resonant mode of a crystalline solid in thermodynamic equilibrium is directly measured, assuming that quantum effects can be neglected it coincides with temperature except for a proportionality factor. This is due to the principle of energy equipartition and the equilibrium hypothesis. However, most natural systems found in nature are not in thermodynamic equilibrium and thus the principle cannot be granted. We measured the extent to which the low-frequency modes of vibration of a solid can defy energy equipartition, in presence of a steady state heat flux, even close to equilibrium. We found, experimentally and numerically, that the energy separately associated with low frequency normal modes  strongly depends on the heat flux, and decouples sensibly from temperature. A 4\% in the relative temperature difference across the object around room temperature suffices to excite two modes of a macroscopic oscillator, as if they were at equilibrium, respectively, at temperatures about 20$\%$ and a factor 3.5 higher. We interpret the result in terms of new flux-mediated correlations between modes in the nonequilibrium state, which are absent at equilibrium.
\end{abstract}

\pacs{05.40.-a,05.70.Ln,02.70.Ns}

\maketitle

The Equipartition of energy relates the microscopic motions to macroscopic observables, when quantum effects are negligible~\cite{tolman}. For systems in thermodynamic equilibrium, the principle states that every quadratic degree of freedom within the Hamiltonian possesses on average an energy $k_B T/2$ , with $k_B$  the Boltzmann constant and $T$ the absolute temperature. Outside equilibrium, such energy content becomes unknown. Nonequilibrium situations are nevertheless the most common in  nature. A simple nonequilibrium situation is produced by temperature differences~\cite{lebon,evans2008}. For fluid systems it has been argued theoretically~\cite{kirkpatrick,ronis,dezarate} and demonstrated experimentally that spatio-temporal fluctuations grow markedly for the lowest wave numbers~\cite{law,li1994,dorfman,li1998,vailati}. Nonequilibrium stochastic processes~\cite{bertini} and nonequilibrium molecular dynamics techniques~\cite{searles2007} have led to similar conclusions, with a growth of the correlations as the signature of the departure from equilibrium. In Extended Thermodynamics~\cite{jou1997}, either local thermal equilibrium is assumed, or the temperature is defined via an extended flux-dependent entropy~\cite{casasvazquez2003}. In disordered systems, the notion of time-scale-dependent  effective temperatures is often used~\cite{cugliandolo}. Solids are seldom investigated in this context, perhaps because their microscopic degrees of freedom are generally assumed to satisfy local equilibrium. A more prolific enterprise is constituted by research in small and low-dimensional systems~\cite{fransson,lepri2003,dhar,miller,rieder,kato,kannan}.

We report an experimental study of the mean energy of harmonic oscillators in and out of equilibrium. The harmonic oscillator is ubiquitous and is applied to the most diverse situations in both classical and quantum physics~\cite{moshinsky}. In the low losses approximation, the response of a system is often written in terms of normal modes, each considered as an independent harmonic damped oscillator. Our oscillators are low-frequency acoustic modes of vibration of a macroscopic aluminum piece~\cite{conti2010}. The nonequilibrium steady states (NESS) are due to sustained thermal differences across the piece.

\section{Oscillator experiment}
\begin{figure}
\centering{
\includegraphics[width=13.0cm,clip=,angle=0]{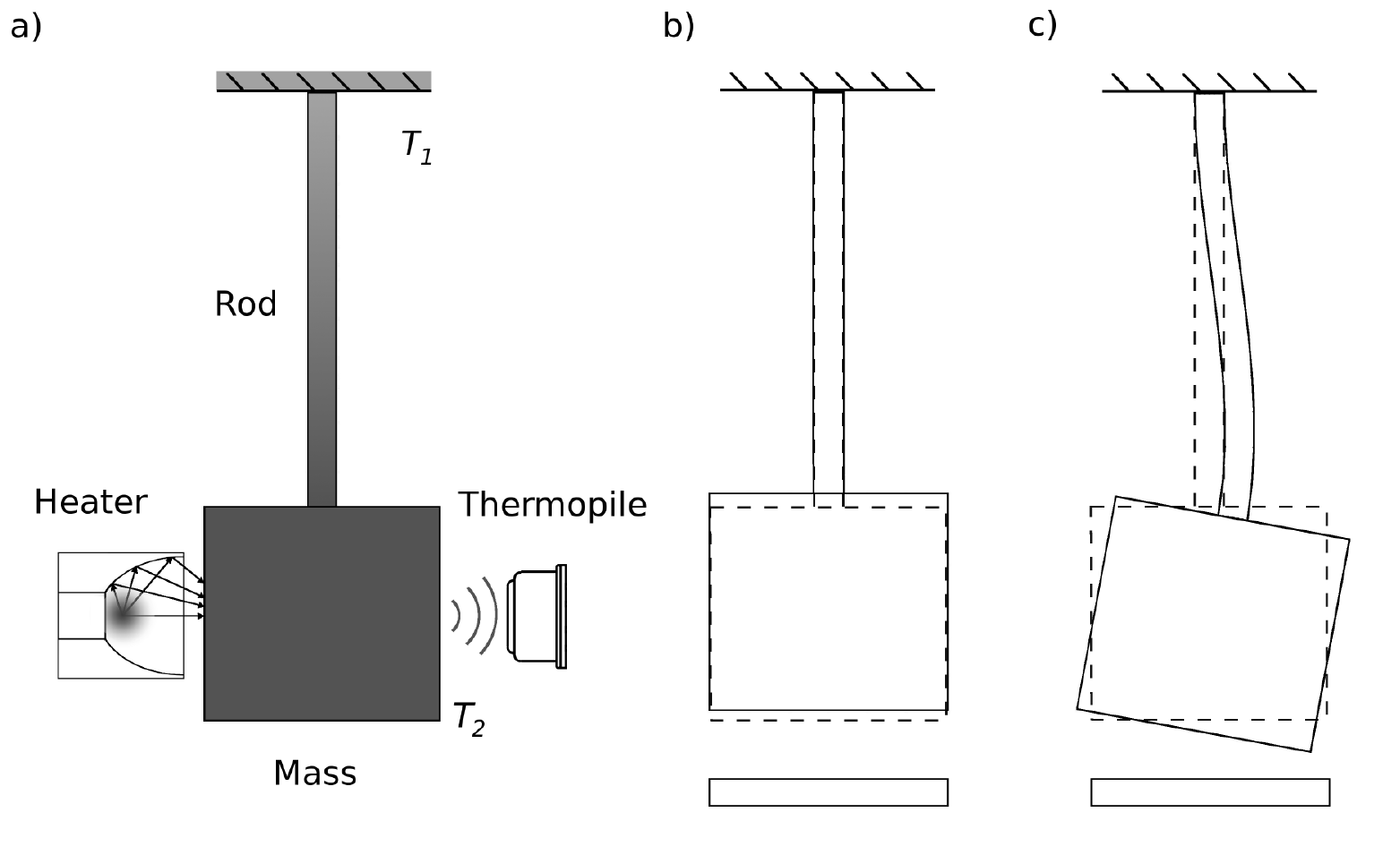}}
\caption{\label{fig:modes} Schematic drawing of the elastic body consisting of a rod with one extreme fixed and the other loaded by a mass and free to move. A pair of thermometers sensed the temperatures $T_1$ and $T_2$ of the rod extremes (a): for the latter case a contact-less thermopile was used. The temperature $T_1$ of the rod top end was stabilized actively; when the heat source shown on the left of the bottom mass was switched on, a thermal difference was set across the body. The thermal profile along the rod in the steady state is shown in gray code (increasing temperature from light to dark). The central (b) and right (c) figures show the deformation of the body respectively in the transverse and longitudinal mode. The shape of the elastic body at rest is shown as dashed line for comparison. The end mass vibrations were measured by a capacitive readout formed by facing the bottom surface of the mass to a fixed and electrically insulated electrode, shown as a rectangle in the parts (b) and (c).}
\end{figure}

The experimental setup is described fully in Ref.~\cite{karapetyan}: here, and in the supplementary material, we add a few more details. To
summarize, we hosted 3 resonant devices in a vacuum environment: they were isolated from
external perturbations via a cascade of mechanical filters so that the dominant noise force
acting on them was that of thermal origin, at least in the frequency range where the lowest
transverse and longitudinal acoustic mode resonated. Each resonator consisted essentially of
a squared cross-setion, 0.1~m long rod hold parallel to the vertical and whose top end is clamped and the bottom end is loaded
by a cuboid mass and is free to move (see also Fig.~\ref{fig:modes}). Each cuboid mass was about 0.2~kg heavy: they were all different so to avoid coupling between the resonators.  Parallel to the vertical rod, on opposite
sides, were two protrusions that were used to realize a capacitive readout. These parts were
machined from a single piece of an aluminium alloy, namely Al5056: we refer to them as the
payloads and in this work we report the results for one of them.

\begin{figure}
\centering{
\includegraphics[width=13.5cm,clip=,angle=0]{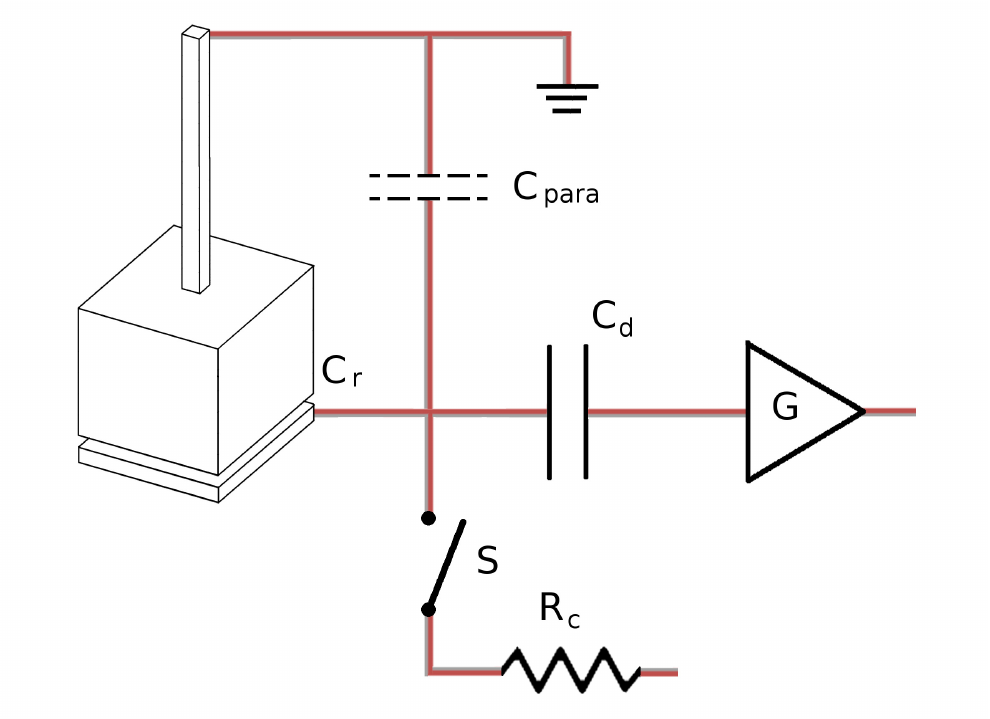}}
\caption{\label{fig:circuit}: Schematics of the capacitive readout for measuring the vibrations of the mass at the
rod bottom end. A parallel plate capacitor ($C_r$ ~160~pF) was formed by the bottom surface of
the mass at the rod bottom end and a fixed plate; the capacitor was biased to a constant dc
voltage through a large resistor ($R_c$) and then disconnected from it by opening the relay S,
which was housed close to the oscillator. Static capacitances in parallel to $C_r$ , represented by
the capacitance $C_{para}$ (~160~pF) in the picture, had the net effect of reducing the sensitivity of
the readout. The signal from the overall parallel capacitor was sent to an amplifier (gain
G~=~100 at the longitudinal resonance) through a $C_d$=100~nF decoupling capacitor.}
\end{figure}

Figure~\ref{fig:modes} shows schematically the deformation of the piece corresponding to the first transverse and longitudinal modes (our two oscillators), which resonate respectively at about 324~Hz and 1420~Hz. Vibrations of the payload bottom mass were measured by a capacitive readout (see Figure~\ref{fig:circuit}) realized between its bottom face and a metal plate (which we name the fixed electrode) supported by the payload protrusions and insulated from them by a couple of PTFE spacers
(100~$\mu$m thickness): the facing surfaces of the bottom mass and plate were finished by diamond turning so to avoid
discharges when applying high electric fields. To tighten the screws that hold the fixed electrode, a calibrated torque was used in order to guarantee a
uniform compression of the PTFE spacers and then the capacitor plates to be parallel. As a result the capacitor $C_r$ sensing the mass vibrations (about 160~pF) was in parallel to the static
capacitance of the PTFE spacers and of the PEEK washers holding the fixed electrode: we
called $C_0$ this capacitance and $C_{washer}$ the parallel capacitance of the PEEK washers.
Additional parallel static capacitances were that of the 0.3~m long wire connecting the fixed
electrode to the decoupling capacitor and that of an additional input line of the amplifier used
for tests and calibrations of the electronics: we called $C_{wt}$ the sum of these two capacitances.

The capacitive readout worked in a constant charge configuration being firstly
connected to a dc high voltage (HV) source (up to 400~V) and then disconnected from it (see
Figure~\ref{fig:circuit}) via a relay housed close to the oscillator: then the capacitance gap changes caused
a voltage signal which was sent to a nearby low-noise band-pass amplifier (core component is
a N-channel junction FET BF862; the gain was set to 100 at the longitudinal resonance) and
then digitized (8~kS/s with antialiasing filter). A low-loss decoupling capacitor (100~nF) was
inserted between oscillator and amplifier in order to use a ground referred
preamplifier (see Figure~\ref{fig:circuit}): in this way after disconnecting the HV source, the dc voltage
across the readout capacitance remained constant even when the latter changed for thermal expansion caused by an applied
thermal difference (see the supplementary material), because of the high value decoupling capacitor. To avoid any risk of
saturation by the low frequency mechanical noise, the amplifier had a negative feedback
loop directly to its input for frequencies up to 200~Hz, thus implementing a second order
high pass filter having a slope of 40~dB/decade. We note that the effect of the static
capacitances in parallel to $C_r$ was that of reducing the sensitivity of the capacitive readout, by
an amount equal to the ratio between the total parallel capacitance (neglecting the decoupling
capacitor) and the capacitor $C_r$ measuring the mass vibrations, $(C_0+C_{wt})/C_r \sim 2$. See the supplementary material for further details.

The Power Spectral Density (PSD) of the amplifier output was computed and time averaged:
for the transverse mode we averaged the PSD from 25 consecutive, non overlapping buffers of 1048576 samples each,
resulting in one average PSD each 54 minutes with a frequency resolution of 0.00763~Hz,
while for the longitudinal mode we averaged over 50 consecutive, non overlapping buffers of 262144 samples each, resulting
in one average PSD each 27 minutes with a frequency resolution of 0.03~Hz. Before
performing the Fast Fourier Transform (FFT), the data were filtered using 4-term Blackmann-Harris window. At low frequencies the noise was dominated by the mechanical noise of the environment; above 200~Hz the mechanical noise became negligible,
being filtered by the suspension system, and the noise level was set by the amplifier: at about
1.3~kHz the corresponding input noise level was $4\times 10^{-9}$~V/$\sqrt{Hz}$. A few
peaks in the PSD were still present: harmonics of the mains and the transverse and
longitudinal modes of the payloads. Above 3~kHz the internal resonances of the suspension
components and of the payload were also evident.

\begin{figure}
\centering{
\includegraphics[width=13.5cm,clip=,angle=0]{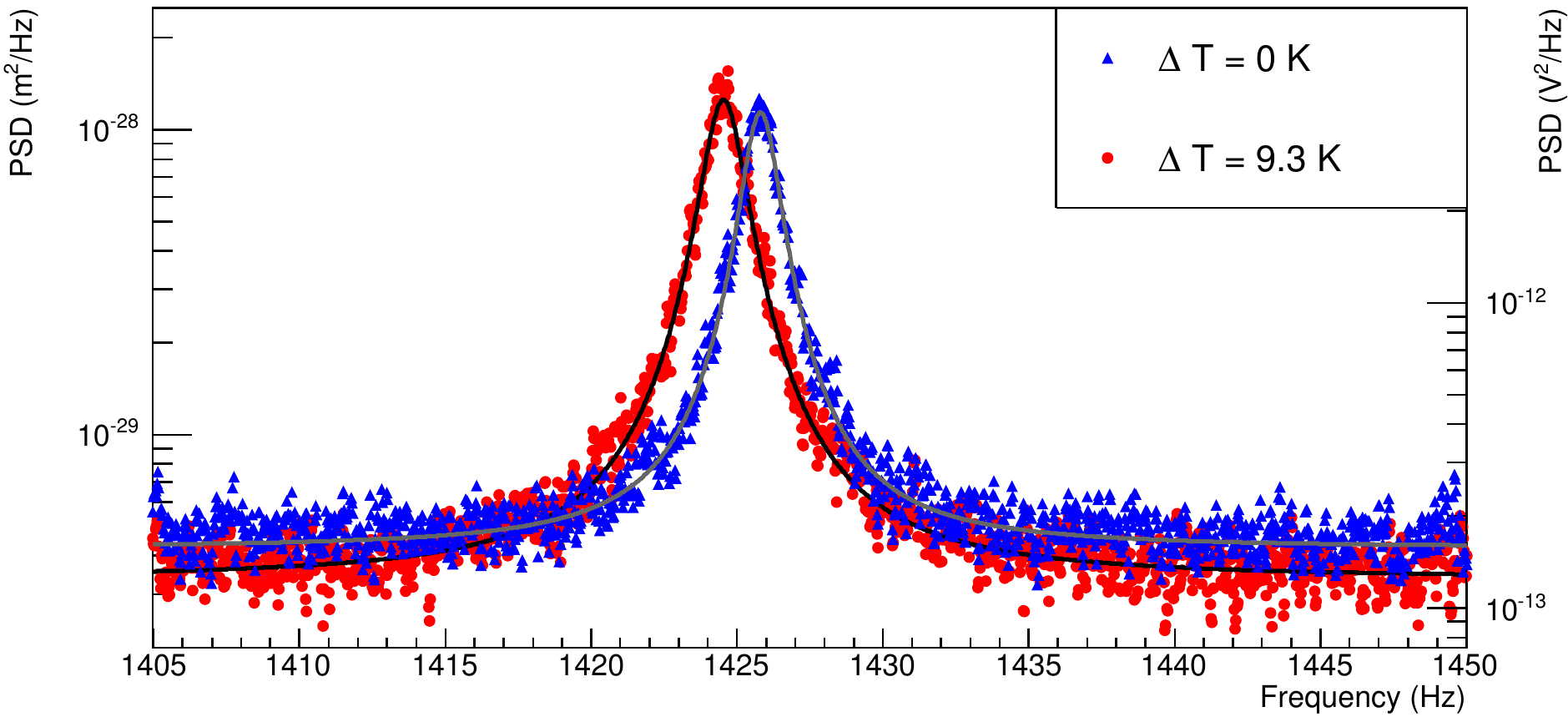}}
\caption{\label{fig:sp} (Color online) Time-averaged Power Spectral Densities (PSD) of the amplifier output around the first longitudinal mode in equilibrium ($T_{avg} = 288.14$~K; blue triangles) and NESS ($\Delta T = 9.3$~K, $T_{avg} = 292$~K; red circles). The PSD are taken with the same sensitivity of the capacitive readout (about $2\times 10^6$~V/m): on the left axis they are converted in terms of mass vibrations. The gray line is the fit of the equilibrium PSD: $T_{eff} = \left[ 319 \pm 5(stat.) \pm 18(syst.)\right]$~K. The black line fits the NESS PSD: $T_{eff} = \left[ 402 \pm 6(stat.) \pm 18(syst.)\right]$~K.
}
\end{figure}

The average PSDs were fitted in a small frequency range around the mode resonances: the fit
region was 20~Hz and 3~Hz respectively for the longitudinal and transverse mode. The fit curve
was:
\begin{equation}
y (f) = \frac{p_1 \, p_2}{4 \left( f-p_0 \right)^2 + p_1^2} + p_3
\end{equation}
where $p_0$ is the resonant frequency, $p_1$ is the Full Width at Half Maximum (FWHM), $p_2$ is $2/\pi$
times the area of the Lorentzian curve and $p_3$ is the constant fitting the noise level out of the
resonance. Figure~\ref{fig:sp} shows a typical average PSD around the resonance of the longitudinal mode with the fitting curve. 

In the case of the longitudinal mode, the detailed knowledge of the experiment and the fact that the electric field is parallel to the mass vibration allow to convert the output signal into longitudinal vibrations of the mass, with now free parameters. Hence the area of the fitting curve becomes an estimate of the mean square vibration $\left\langle x_l(t)^2 \right\rangle $  of the oscillator, where the angular brackets denote a time average. At thermodynamic equilibrium with low losses, in the absence of external noises the law of equipartition dictates that:
 \begin{equation}
 \left\langle x_l(t)^2 \right\rangle = \frac{k_B \, T}{m_l \, \omega_l^2} \label{eq:Teff}
 \end{equation} 
where $m_l$ is the mass of the longitudinal mode (well approximated by the freely moving mass) that resonates at $\omega_l$. Thus, thanks to equipartition, the thermodynamic temperature can be inferred from the energy stored in the mode. In general, i.e. even in NESS, we call effective temperature $T_{eff}$  the quantity computed from the mode energy as in Eq.~(\ref{eq:Teff}).

In parallel with the acquisition of the signals from the 3 payloads, we also acquired at
$f_s=0.1$~Hz the signals of the thermometers measuring the temperature $T_0$ of the aluminium
mass where all the 3 payloads are fixed to, the temperatures $T_1$ of the rod top end and $T_2$ of
the mass at the rod bottom for all the 3 payloads, and a few diagnostics signals. To avoid introducing losses,
contact-less thermopile was used to measure the temperature of the bottom mass. To reduce the electronic noise in the thermopile output, we processed it with a
triangular filter, similarly to what reported in ref.~\cite{conti2012}. We estimate a total error of
$\pm$0.2~K on $T_2$, with a uniform distribution. To set nonequilibium states, the oscillating mass was exposed to a thermal heat source. We obtained both thermal equilibrium states and NESS at different $T_1$  and $T_2$.
 
\begin{figure}
\centering{
\includegraphics[width=13.5cm,clip=,angle=0]{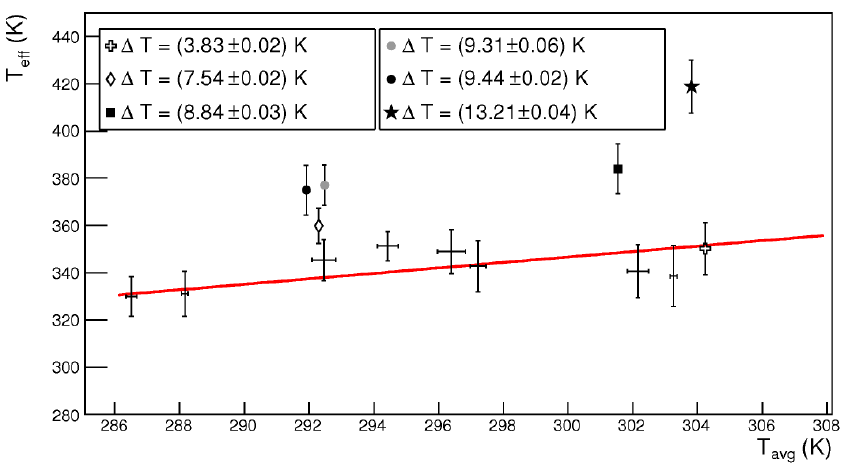}}
\caption{\label{fig:Teff} (Color online) The effective temperature of the first longitudinal acoustic mode of resonance against the mode average temperature; error bars correspond to the rms of the measurements. The dots refer to the equilibrium state (for clarity, the measurements were grouped in 2~K intervals of $T_{avg}$).  The line is the linear fit of all equilibrium measurements. The results in NESS corresponding to different heat flows are shown with different markers, as explained in the legend. The quoted errors in $\Delta T$ represent the rms for the given state; $\Delta T$ values are also affected by a systematic error of $\pm 0.2$~K. The effective temperatures are also affected by a systematic error, equal to 18~K in equilibrium and 20-30~K for the different NESS.}
\end{figure}

Equilibrium measurements were performed in a 25~K interval around room temperature. 
The quantities used to compute $T_{eff}$ from Eq.~(\ref{eq:Teff}) are listed along with their error in Table S1.
We estimated the capacitance of the PEEK washers via numerical methods using Finite
Element Method (FEM). The value of the oscillator mass was set to the mass at the rod end
plus half of the rod mass, with an error which accounts for half of the rod mass, machining
precision and error in the density: this estimate was also confirmed by a numerical analysis
with FEM. The value of the thermal expansion of aluminium and PTFE permittivity were
obtained from the literature. All other quantities were measured independently.
The effective temperatures computed using Eq.~(\ref{eq:Teff}) are plotted in Figure~\ref{fig:Teff} against the average temperature $\left( T_1+T_2\right) /2$  (which reduces to $T_1$ in this case). The effective temperatures are affected by a systematic error of 18~K and by a random error between 4~K and 7~K. The systematic error was estimated both via an analytic error propagation procedure adding the separate contributions in quadrature and via a numerical Montecarlo analysis; the largest source of systamtic error are the measurements of the capacitances at the amplifier input and the estimate of the
oscillator mass. 

The proportionality between $T$ and $T_{eff}$ in equilibrium is shown by the linear fit of the equilibrium data. The line slope results in $1.156\pm0.001(stat.)\pm0.06(syst.)$, indicating that the oscillator was dominated by the thermal noise and that, thanks to equipartition, $T_{eff}$ operates as a virtual thermometer for the oscillator at equilibrium. For simplicity, because of our limited range of temperatures, we shall take the equilibrium effective temperature of the longitudinal mode always equal to the mean value $T_{EQ}$ of all the equilibrium measurements, with random error equal to their root mean square (see Fig. S4 of Supplementary Material): $T_{EQ}=[338 \pm 12$~(stat.)~$\pm 18$~(syst.)$]$~K. Adding the statistical and systematic contributions, the measured $T_{EQ}$ agrees with the thermodynamic temperature $T_{avg}$ within 2 standard deviations: we remark that no free parameters are used in this derivation.

We repeated the same kind of measurements in the presence of heat fluxes. We observed an increase of the PSD around the resonance, as shown in Figure~\ref{fig:sp}. The resulting effective temperatures are plotted in Figure~\ref{fig:Teff} against the mode average temperature $T_{avg}$. For the longitudinal mode this coincides~\cite{conti2012} with $\left( T_1+T_2\right) /2$. The total error consists of a statistical component between 4~K and 7~K and a systematic one between 20~K and 26~K. The largest
systematic error contribution in the nonequilibrium $T_{eff}$ is due to the calibration error in $\Delta T$, the error in
the thermal expansion coefficient $\alpha$ of Al5056 and in the oscillator mass $m_l$. In absolute terms, the effective temperature grows with increasing temperature differences more rapidly than with the temperature, as can been seen in Figure~\ref{fig:Teff}. 

\begin{figure}
\centering{
\includegraphics[width=13.5cm,clip=,angle=0]{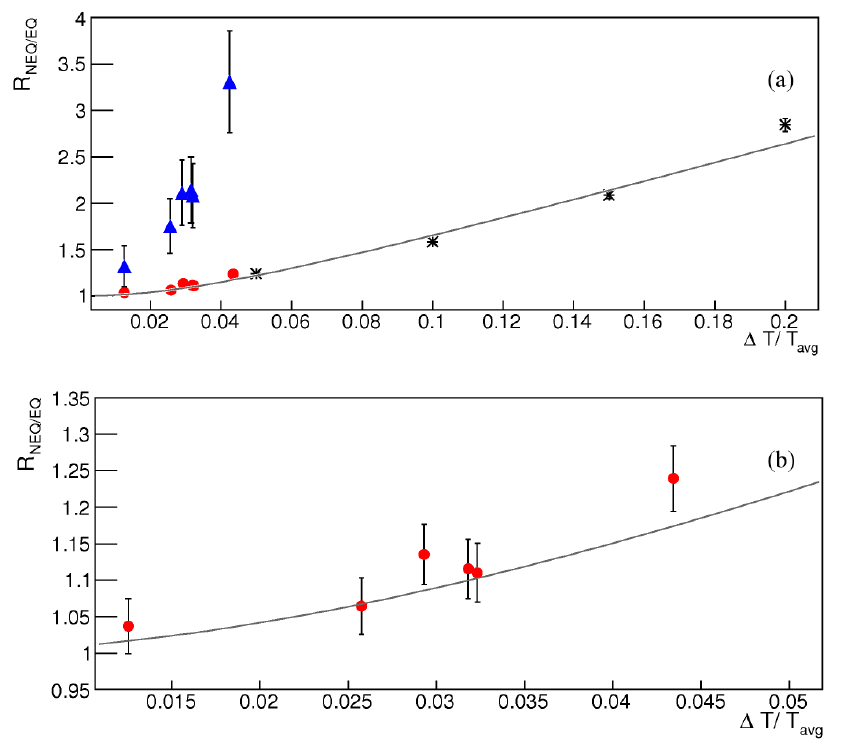}}
\caption{\label{fig:ratio} (Color online) (Part a) Ratio $R_{NEQ/EQ}$ of the average effective temperature of the transverse (blue triangles) and longitudinal (red circles) acoustic modes in NESS over their average effective temperature in equilibrium, against the normalized thermal difference $\left( T_2 - T_1 \right)  /T_{avg}$. The error bars show the statistical uncertainty. Along the horizontal axis they are smaller than the size of the points. The black stars show the results of the numerical experiment with their error bars. The gray line is the best fit of the numerical data with the function of Eq.~(\ref{eq:x2neq}), resulting in $a= 108 \pm 1$ ($a$ defined in main text, following Eq.~(\ref{eq:asymptotic2})). Part b of the figure shows the results of the longitudinal mode and the line fitting the numerical result.
}
\end{figure}

For a better comparison between equilibrium and nonequilibrium states, we computed the ratios $R_{NEQ/EQ}=T_{eff}/T_{EQ}$; they are plotted in Fig.~\ref{fig:ratio} against the temperature difference at the rod extremes normalized by $T_{avg}$. Error bars represent the statistical uncertainty: we estimated a systematic error of $\pm 0.01 \div 0.03$ along the vertical axis and of $\pm 7\times 10^{-4}$ along the horizontal axis. Table S2 of the supplemantary materia lists the values of $R_{NEQ/EQ}$ along with their statistical and systematic errors. We note that in the temperature ratio a few quantities cancel out: this is the
case for instance of the amplifier gain $G$ and of the oscillator mass $m_l$. Thus the temperature
ratio is independent of these quantities and hence of their errors. We note that $R_{NEQ/EQ}>1$. For instance, at the maximum relative temperature difference, $R_{NEQ/EQ}>1$ by more than 4 standard deviations: a $4\%$ relative temperature difference is enough to increase the nonequilibrium $T_{eff}$ by 20$\%$ ie to raise it above the hottest physical temperature present in the piece. Thus away from, even if close to, equilibrium, $T_{eff}$ is not a valid measure of the physical temperature.

For the same payload we observed the transverse mode depicted in Fig.~\ref{fig:modes}-a: the rod bends and the bottom mass performs a rotation. In the corresponding NESS, $T_{avg}$ is not simply $\left( T_1+T_2\right) /2$ (see Figure S3 of Supplementary material). Therefore we estimated the mode average temperature $T_{avg}$ from the estimated mode resonant frequency, with an error of 1~K. We computed the ratios between the effective temperatures in the NESS (see the Supplementary Material) and in equilibrium, which are plotted in Figure~\ref{fig:ratio} with their statistical uncertainty: we estimated the systematic errors along the vertical axis (reported in Table S2 of Supplementary material) and a total error of $\pm 7\times 10^{-4}$ along the horizontal axis. A relative temperature difference of at most 4\% caused an up to threefold increase of $T_{eff}$. In a NESS, this would imply an effective temperature of around $900$~K, with the hot section still at about $310$~K. Again, adding the statistical and systematic contributions in quadrature, at the maximum relative temperature difference, $R_{NEQ/EQ}>1$ by more than 4 standard deviations.

\section{Numerical experiment}
\begin{figure}
\centering{
\includegraphics[width=13.5cm,clip=,angle=0]{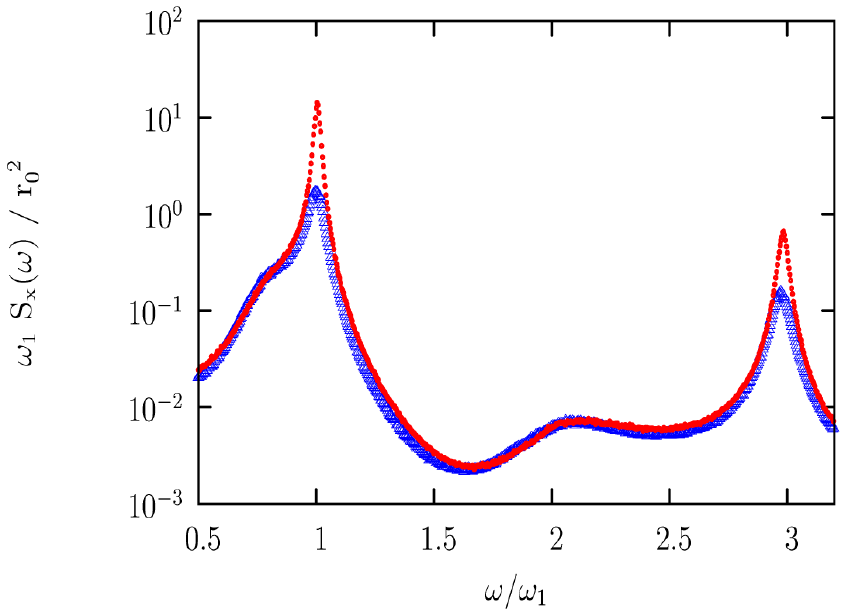}}
\caption{\label{fig:sp_numeric} (Color online) The dimensionless quantity $\omega_1 \, S_x\left(  \omega \right) /r_0^2$, as a function of the dimensionless quantity $\omega / \omega_1$, where $S_x \left(  \omega \right)$ is the average PSD of the position of the last particle of the one-dimensional Lennard-Jones chain. $\omega_1$ is the resonance frequency of the first mode of vibration and $r_0$ is a measure of the inter-particle average distance, as defined by the Lennard-Jones potential. The peaks are the first and second mode. The blue open triangles are equilibrium simulations at a given temperature $T$, the red circles are nonequilibrium simulations with relative temperature difference $T_2-T_1=0.2 T$, around the same $T$.
}
\end{figure}

For a numerical investigation, we employed a one-dimensional model~\cite{conti2012} in which identical particles interact with their first- and second-neighbors, via Lennard-Jones type potentials. One end of the chain was clamped, and its immediate neighbors were thermostated at $T_1$; the other end was free~\cite{degregorio2011}, and the last two particles were thermostated at $T_2$. The temperatures
at each side of the chain were controlled via the Nos\'e-Hoover algorithm, determining locally
the average kinetic energy. Thus, differently from the experiment, here we have two
thermostats at the two ends, instead of one thermostat and an over-imposed heat flux. We investigated numerically the time evolution $x(t)$ of the total length of the chain, after equilibrium ($T_1 = T_2$) or NESS ($T_1 < T_2$) was reached. $T_2-T_1$ was varied with $\left( T_1 + T_2\right) /2$ constant. For
each prescribed pair of $T_1$ and $T_2$, 1600 contiguous time intervals were selected from the
stationary states of 20 distinct molecular dynamics simulations. In each interval the FFT was calculated and then multiplied by its complex conjugate, then the
average over all the time intervals is performed. The (averaged) PSD is shown in Figure \ref{fig:sp_numeric} in the frequency range around the first two resonances; the figure shows the
results for both the equilibrium and nonequilibrium case.

From Fourier analysis we identified the successive longitudinal modes of vibration of the chain. We observed a systematic increase of the area of the Fourier components around the first modes with the magnitude of the temperature differences. The effect is most pronounced for the first mode and fades away for the highest frequency modes we analyzed. In Figure~\ref{fig:sp_numeric} we show the (averaged) PSD of the variable $x(t)$ encompassing the first two modes in equilibrium and nonequilibrium (when the extremal temperatures differ by 20~$\%$): the comparison evidences a behavior similar to the experimentally observed one, cf. Figure~\ref{fig:sp}.

In Figure~\ref{fig:ratio}-(a), we show the estimate of the area delimited by the Fourier spectrum around the first mode in NESS, normalized by the equilibrium value at $\left( T_1 + T_2 \right) /2$ (the starred points). For each $\Delta T/T$, the uncertainty in the estimation of $R_{NEQ/EQ}$ originates from the uncertainty in the interval where to integrate the first peak of the equilibrium and nonequilibrium PSD’s: namely it is the half-interval of the variations of $R_{NEQ/EQ}$ when using different integration ranges. The maximum relative error is 1~$\%$. The quantitative consistency with the experimental results  suggests that $R_{NEQ/EQ}$ for the first longitudinal mode is a universal function of $(T, \, \Delta T)$, irrespective of the microscopic details. 

\begin{figure}
\centering{
\includegraphics[angle=90,width=13.5cm,clip=]{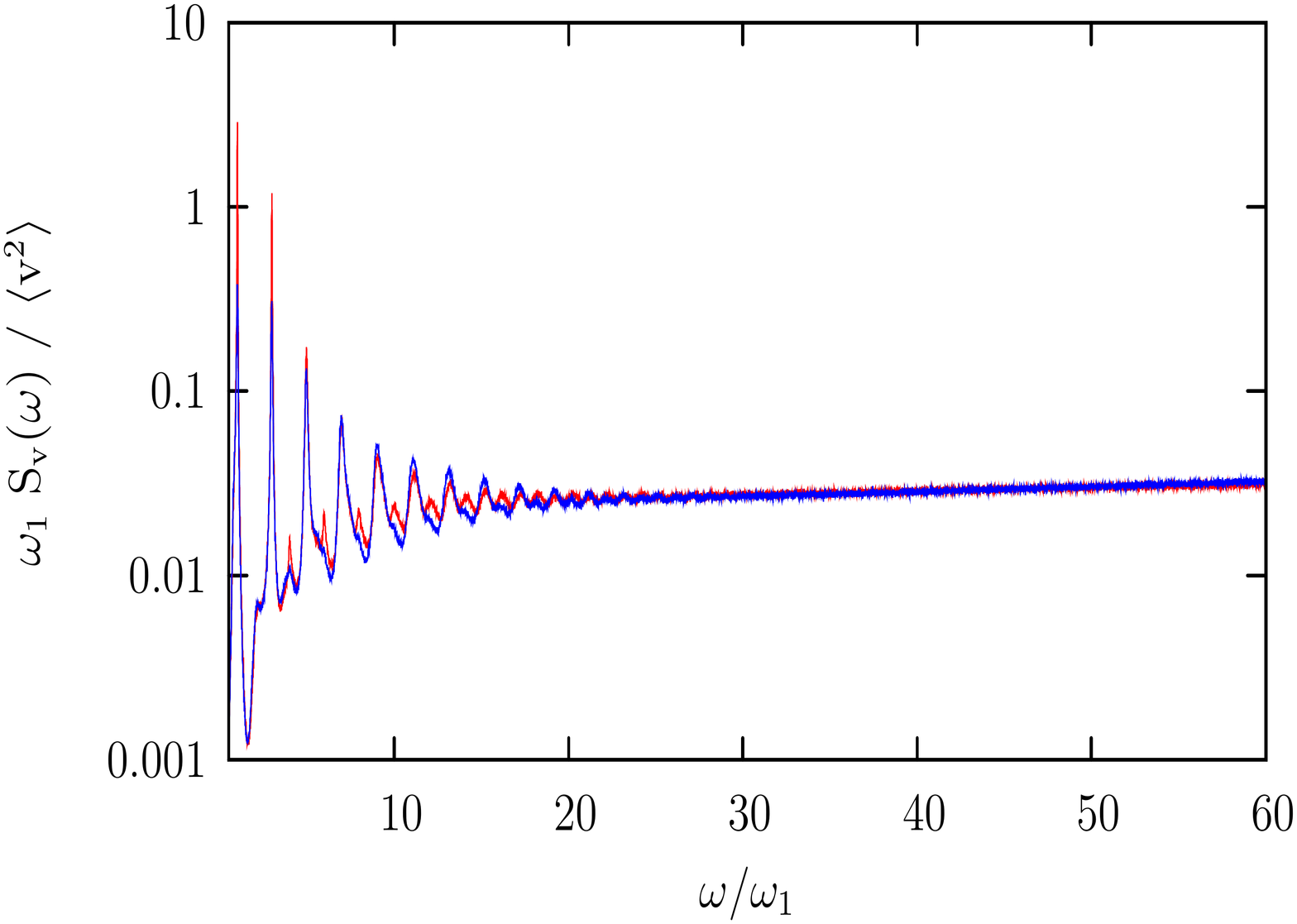}}
\caption{\label{fig:sp_numeric_v} (Color online) The dimensionless quantity $\omega_1 \, S_v\left(  \omega \right) / \langle v^2 \rangle$, as a function of the dimensionless quantity  $\omega / \omega_1$, where $S_v\left(  \omega \right)$  is the average PSD of the velocity of the last particle of the one-dimensional Lennard-Jones chain.  $\langle v^2 \rangle$ is the time-average of the last particle's velocity squared (which is proportional to $k_B T_2$). The set of simulation settings from which these data are taken is the same as that of Figure~\ref{fig:sp_numeric}, except that now we consider velocities rather than positions. Hence a larger range is shown on the horizontal axis. The color code is the same as in Fig.~\ref{fig:sp_numeric}.
}
\end{figure}

For discussing the simulation results in a wider frequency range, i.e. extending much above the first few resonances, we also consider the Power Spectral Density of the velocity variable $v(t)$, which is shown in Figure~\ref{fig:sp_numeric_v}. This high-frequency regime is consistent with the constraint that the full integral of the velocity spectra is proportional to the kinetic energy, which is set by the thermostats to be proportional to the target temperature $T_2$. It is thus consistent with the intensity of the nonequilibrium spectrum being a fraction higher than the equilibrium one, and that a local law of equipartition applies to the fastest oscillations.

\section{Theoretical model}
For a theoretical interpretation concerning the longitudinal modes, assume that the equilibrium probability distribution of the microscopic states for a low losses system is approximated by the canonical distribution for a chain of $N$ harmonic oscillators: 
\begin{equation}
P_{EQ}({\bf x},{\bf v})=\frac{e^{-H({\bf x},{\bf v})/k_BT}}{Z} \label{eq:Peq}
\end{equation}
\begin{equation}
H=\frac{1}{2} \sum_i \mu_i(\omega_i^2 x_i^2+ v_i^2 ); \quad Z= \int {\rm d}^Nx ~ {\rm d}^N v ~ P({\bf x},{\bf v})  \nonumber
\end{equation}
The $x_i$'s and $v_i$'s here represent the (approximate) projection of the full $N$-dimensional dynamics over the $i$-th mode of oscillation, which resonates at $\omega_i$, $\mu_i$'s are the reduced mass coefficients. In the case in which nearly all the mass $M$ of the chain belongs to the last particle, a situation which emulates the experimental setup, $\omega_1 \simeq \omega_l$ and $\mu_1$ is well approximated by $M$, and by $m_l$ in turn. Thus, Equation (\ref{eq:Peq}) implies Equation (\ref{eq:Teff}).

The NESS heat flux is commonly defined via cross-terms  $x_i \, v_j$ ($i\neq j$), implying that the emergence of correlations between modes absent in equilibrium parallels the onset of heat flows. In one-dimensional models
under the harmonic approximation~\cite{lepri2003,dhar,miller,rieder,kato}, this leads to the following
expression for the NESS heat rate:
\begin{equation}
J= -\frac{1}{2N}\sum_{i , k}^{1,N} j_{ik}(x_iv_k-x_kv_i) = -\frac{1}{N}\sum_{i \neq k}^{1,N} j_{ik} \, x_iv_k \label{eq:Jtot}
\end{equation}
with $j_{ik}$ a coupling constant between the modes $i$ and $j$. In fact, without loss of generality, in a NESS we can assume the probability distribution to be invariant under under the change $x_i \, v_k \rightarrow -x_k \, v_i$ , because the average quantity $\langle x_i x_k \rangle$  has null time derivative. This implies the consistency of assuming each one of the expressions of Equation~(\ref{eq:Jtot}). 

In terms of probability distributions, a very simple form $P_{NEQ}({\bf x},{\bf v})$ yielding nonzero expectation values for $x_i \, v_j$ was obtained from the equilibrium canonical distribution Eq. (\ref{eq:Peq}), with the substitution $H \longrightarrow H + \gamma \, k_B T J$~\cite{miller,kato}:
\begin{equation}
P_{NEQ}({\bf x},{\bf v})=\frac{e^{-H({\bf x},{\bf v})/k_BT}-\gamma J({\bf x},{\bf v})}{K} \label{eq:Pneq_gen}
\end{equation} 
with $K$ the new normalization factor which substitutes $Z$. Notice that for the probability distribution of Eq.~(\ref{eq:Pneq_gen}) to be well defined, one must also assume that $\gamma$ has an upper bound, differently from the inverse temperature $(k_BT)^{-1}$. Equation~(\ref{eq:Pneq_gen}) is equivalent to the hypothesis that  is exponentially quadratic and marginally Gaussian in all its variables, and therefore a multivariate Gaussian distribution.

The requirement that $P_{NEQ}$  satisfies Equation~(\ref{eq:Pneq_gen}), with $J$ defined as in Eq.~(\ref{eq:Jtot}), is less general than what one has to assume for the analysis of a single mode of vibration under a constant heat flux. 
 Here, we restrict our attention to a single mode, positing that
\begin{equation}
P_{NEQ}(x_1,v)=\frac{e^{-M \omega_1^2 x_1^2/ 2 k_BT + \lambda M \omega_1^2 x_1 v - \mu  v^2/2 k_B T} }{\kappa} \label{eq:Pneq}
\end{equation}
where $v$ is a (effective) velocity variable correlated with $x_1$, $\mu$ its reduced mass, $P(x_1,v)$ the joint probability distribution of $x_1$ and $v$, $\lambda$ a parameter in units of inverse heat rate. $\kappa$ is the normalizing factor, whose explicit expression can be derived provided $\lambda^2 M\omega_1^2 < \mu/(k_BT)^2$, and via appropriate derivations it leads to expressions for $\langle x_1 v \rangle_{NEQ} $ and $\langle x_1^2 \rangle_{NEQ}$. 
Notice that Equation~(\ref{eq:Pneq}) can be derived from successive integration of Eq.~(\ref{eq:Pneq_gen}) over the mode variables correlated with $x_1$. Alternatively, one may conjecture that the first mode is primarily correlated with only one mode out of the remaining $N-1$ degrees of freedom, and that $v$ and $\mu$ are the velocity and mass of that mode. However, we do not advocate that these are the only possibilities leading to the assumption made in Equation~(\ref{eq:Pneq}). Our other key assumption is that, to a good approximation, everywhere in our equations $T$ is a measure of the average temperature of the rod.

By simple integration of Equation~(\ref{eq:Pneq}),
\begin{equation}
\kappa = \frac{2 \pi}{\sqrt{M \omega_1^2 \left[ \mu /\left( k_B T \right)^2 - \lambda^2 M \omega_1^2 \right]}} \, ; 
\, \lambda^2 M \omega_1^2 < \mu /\left( k_B T \right)^2
\end{equation} 	
All second-order moments from Eq.~(\ref{eq:Pneq}) can be derived using combinations of $\kappa^{-1} \left(\partial \kappa/\partial y \right)$ , with $y=\left( k_B T \right)^{-1}$  and $y = \lambda$ alternatively. The average quantities $\langle x_1^2 \rangle_{NEQ}$  and  $M \omega_1^2 \langle x_1 v \rangle_{NEQ}$ therefore are:
\begin{equation}
\langle x_1 v \rangle_{NEQ} = \frac{\lambda}{\mu/\left( k_B T \right)^2 - \lambda^2 M \omega_1^2} \, , 
\end{equation}
\begin{equation}
\langle x_1^2 \rangle_{NEQ} = \frac{\mu}{\lambda M \omega_1^2 k_B T} \langle x_1 v \rangle
\end{equation}
If the fraction of heat rate carried on average by the first mode is defined as $\phi=-M\omega_1^2 \langle x_1 v \rangle$, one obtains:
\begin{equation}
\langle x_1^2 \rangle _{NEQ} =   \frac{\eta }    {\eta - \lambda(\phi)^2} \, \langle x_1^2 \rangle_{EQ} \label{eq:x2neq}
\end{equation}
\begin{equation}
\lambda(\phi)=\frac{1}{2 \phi} (1-\sqrt{1+4 \eta \phi^2}); \quad \eta = \frac{\mu}{M \omega_1^2 (k_BT)^2}  \label{eq:lambdaeta}
\end{equation}
where $\langle x_1^2 \rangle_{EQ}$ is the equilibrium average at the same $T$ occurring in $P_{NEQ}(x_1,v)$ in Eq.~(\ref{eq:Pneq}). We assume that $T$ lies in the range $[T_1,T_2]$, e.g. equal to $T_{avg}$ up to easily controllable corrections, incidentally allowing the physical temperature to be decoupled from the kinetic energy stored in the mode. Asymptotically we have:
\begin{equation}
\langle x_1^2 \rangle \simeq \langle x_1^2 \rangle_{EQ} \, (1+ \eta \, \phi^2); \qquad |\phi| \ll 1/\sqrt{\eta} \label{eq:asymptotic}
\end{equation}
\begin{equation}
\langle x_1^2 \rangle \simeq \langle x_1^2 \rangle_{EQ} \,  \sqrt{\eta} \, |\phi|; \qquad \qquad |\phi| \gg 1/\sqrt{\eta} \label{eq:asymptotic2}
\end{equation}
For $T$ fixed, letting $\phi=c \, \Delta{T}/T$ and $a=\eta \, c^2$,  Eqs.~(\ref{eq:x2neq},~\ref{eq:lambdaeta}) can be rewritten in the dimensionless terms:
\begin{equation}
\zeta (x) = \frac{1}{2 x} \left( 1-\sqrt{1+4 a x^2} \right)
\end{equation}
\begin{equation}
R_{NEQ/EQ}= \frac{a}{a - \zeta(x)^2} \quad ; \quad x=\frac{\Delta T}{T}
\end{equation}
which is used to fit the numerical data of Fig.~\ref{fig:ratio}. Thus, the experimental setup may be deemed to be a protocol to infer the value of the Lagrange multiplier $\lambda$.

For the lowest relative temperature difference:
 \begin{equation}
\left\langle x_1^2 \right\rangle _{NEQ} - \left\langle x_1^2 \right\rangle _{EQ} \propto \Delta T^2
 \end{equation}
which leads to $R_{NEQ/EQ}-1 \propto \left( \Delta T/T \right) ^2$ and which in turn resembles results obtained in different contexts~\cite{dezarate,bertini,casasvazquez2003,crisanti,evans2005}. The experimental data presented in Figure~\ref{fig:ratio} are consistent with this behavour.

\section{Conclusions}
In this work we treated some acoustic modes of a solid as individual observables. Their energy fluctuations increase alongside growing heat fluxes, disqualifying them as effective thermometers. This behavior differs from that of the modes resonance frequencies, which follow the average temperature. We interpreted this result in terms of correlations between normal modes variables in NESS. Systems defying equipartition laws have been discussed for fluids and disordered systems~\cite{morriss1999,baranyai} and crystal models with alternating masses~\cite{miller,kannan}. Our result suggests that the term correcting the equilibrium expectation value should be carefully measured, in cases where high precision is required, already at moderate temperature differences. Although normal mode decomposition is commonly assumed, temperature immediately ceases to be the sole parameter dictating the fluctuations of the delocalized modes.

\ack{We acknowledge the contribution of the European Research Council under the European Community's Seventh Framework Programme (FP7/2007-2013)/ERC grant agreement no. 202680.}


\section*{References}

\bibliographystyle{iopart-num}
\bibliography{noequivibrajsm}



\section*{Supplementary material}

# Effects of breaking vibrational energy equipartition on measurements of temperature in macroscopic oscillators subject to heat flux

## Supplementary Materials

Details of the experimental setup, data acquisition and fitting procedure

The experimental setup is described fully in Ref.(26) and in the main text: here we add a few more details. We call A, B and C the three payloads; in this work we report results for payload C.

In spite of the fact that we implemented a constant-charge readout, a charge leakage was unavoidable, through the PTFE spacers and stray resistances: to quantify it we observed the areas of the curve fitting the average PSD around the longitudinal mode versus time. An exponential fit over periods of steady state (either with applied thermal difference or not) provided estimates of the discharge time constant: the smallest time constant was $3.4 \times 10^6$ sec. In the computation resulting in fig.s 3-5, the sensitivity of the readout was estimated also by correcting the bias voltage according to the discharge in time after the capacitor was biased. We further estimated the effect of the discharge and computed the noise force acting on the oscillating mass due to the discharge current. The shortest time constant corresponded to a discharge current of $9 \times 10^{-12}$ A: under the hypothesis of shot noise (hence $1.5 \times 10^{-15}$ A/√Hz), the latter results into a noise force 5 orders of magnitude less than the thermal noise force of the oscillator and less even than the contribution of the amplifier current noise, described below.

When a thermal difference $\Delta T = T_2 - T_1$ was present, the gap of the readout capacitor shrank because the oscillating mass and the rod expanded:

$$g\big|_{\Delta T} = g\big|_{\Delta T = 0K} - \alpha \, \Delta T \left(\frac{L}{2} + h\right)$$

where $g|_{\Delta T=0K}$ is the gap in equilibrium (90μm for payload C), α is the thermal expansion coefficient of Al5056 (see Table S1), $L$=0.1m is the length of the rod and $h$=0.041m is the height of the cuboid mass. The factor ½ in the expansion of the rod accounts for the fact that a temperature difference ΔT is present between its extremes and is a consequence of the Fourier law. The change of the gap had 2 consequences: an increase of the electric field across the capacitor and a reduction of the total noise of the amplifier as seen at its output, a feature evident around the longitudinal mode resonance, as shown in Figure 3. In fact, the total noise at the output $S_{V,tot}$ was the sum of the additive voltage noise (estimated to be $\sqrt{S_{VV}} \simeq 1.5 \times 10^{-9}$V/√Hz) and the current noise (estimated to be $\sqrt{S_{II}} \simeq 1 \times 10^{-14}$A/√Hz) flowing back in the readout capacitance $C_{tot}$ and then multiplied by the amplifier gain $G$. In formulas:

$$S_{V,tot}(\omega) = G^2 S_{VV} + G^2 \frac{S_{II}}{\omega^2 C_{tot}^2}$$

where the total capacitance depended on the gap changes:

$$C_{tot}(\Delta T) = C_{para} + C_r|_{\Delta T=0K} \frac{g|_{\Delta T=0K}}{g|_{\Delta T}}$$

where $C_{para}$ is the sum of all static capacitances in parallel to the readout one $C_r$ except for the decoupling capacitor (i.e. PTFE spacers, $C_s$, $C_{wt}$) (see Figure 2 of the main text). With the experimental figures of payload C listed in Table S1, around the longitudinal mode resonance the dominant contribution was that of the current noise: we estimated at equilibrium a total noise $S_{V,tot}$ (2π 1422Hz) = 1.6x10$^{-13}$V$^2$/Hz and at $\Delta T$ =12K $S_{V,tot}$ (2π 1422Hz) = 1.2x10$^{-13}$V$^2$/Hz. These estimates (and their ratio) are similar to the experimental findings shown in Figure S1 and Figure 3. The amplifier current noise produced also a force noise across the sensor capacitor, which was more than 3 orders of magnitude smaller than the thermal noise

force of the oscillator. For frequencies close to the transverse mode resonance, the effect on $S_{V,tot}$ was less noticeable (see Figure S2), given the larger additive voltage noise of the amplifier.

Average temperature of the transverse mode

Figure S3 shows the (square of the) resonant frequency of the transverse acoustic mode as function of $(T_1+T_2)/2$, for the equilibrium and nonequilibrium states. We fitted the equilibrium data with a straight line: hence we computed the nonequilibrium transverse mode average temperature as the one that would result in the same resonant frequency if the oscillator were in equilibrium.

With this procedure we end up with a 1K error in $T_{avg}$ for the transverse mode.


Data selection

All data reported in this paper passed a filtering procedure, essentially based on the goodness of the fit (item i and ii in the following list) and on the requirement of the stability of the steady-state (item iv in the following list). We also noticed that the payload A suffered periods of high instability (mean energy raising by up to a factor 100 for several days) of unexplained origin: at the extreme energies, we observed the emergence of a correlation in the transverse mode energy between payload A and C. Thus for this mode we also adopted a selection criterion (item iii in the following list) based on the energy of the payload A. In details the criteria that the data had to meet simultaneously were:

   i.   relative error in the estimate of the resonant frequency $\sigma_{p0}/p_0<10^{-5}$ (for longitudinal mode), $\sigma_{p0}/p_0<2\times10^{-5}$ (for transverse mode) ;

   ii.  relative error in the estimate of the Lorentzian curve area $0<\sigma_{p2}/p_2<0.5$;

iii. steady state, as detailed in the following;

iv. area of the payload A transverse mode $p_2<0.0018V^2$ (for transverse mode only)

To select the measurements of the oscillator energy taken in steady states we put an upper limit to the maximum total time derivative of $T_1$ and $T_2$ during the averaging time of the PSD. Therefore during the averaging time of all measurements of the oscillator energy reported in this paper, the temperatures $T_1$ and $T_2$ satisfied the criterion:

$$\max\left\{\frac{f_s}{15}\sqrt{\left(\frac{T_1(100\times i)-T_1(100\times i+30)}{T_1(100\times i)+T_1(100\times i+30)}\right)^2+\left(\frac{T_2(100\times i)-T_2(100\times i+30)}{T_2(100\times i)+T_2(100\times i+30)}\right)^2}\right\}<6\times10^{-8}\,s^{-1}$$

where $f_s$ is the sampling frequency of the thermometers and $i$ runs over the samples collected during the averaging time of the PSD. This selection was applied to data taken both when the thermal heat source was on and off. For $T_1=T_2=300K$ this corresponds to maximum time derivative of $9\times10^{-6}$K/s during each PSD averaging time.

After cuts i, ii and iii were applied, cut iv further rejected about 20% of the data of the transverse mode. Summing up, in this paper we presents 1599 and 4412 measurements respectively for the transverse and longitudinal mode: they correspond to a total observation time of 61 days and 85 days respectively for the transverse and for the longitudinal mode.

Determination of the effective temperature and on error estimations

The quantities used to compute $T_{eff}$ from Eq. (2) are listed along with their error in Table S1. If the dominant noise source is the thermal one, one would expect to observe a linear dependence of $T_{eff}$ with the thermodynamic temperature for the equilibrium data shown in Figure 4. However, the range where we varied the thermodynamic temperature was limited to about twice the random error of $T_{eff}$. Thus in this work we have considered the equilibrium effective temperature of the longitudinal mode to be equal to the mean value $T_{EQ}$ of all the

equilibrium measurements, with an error equal to their root mean square. From the histogram (shown in Fig.S4) of all the effective temperatures measured in equilibrium we obtain $T_{EQ}=$ [338±12(stat.) ±18(syst.)]K.

The systematic error in $T_{eff}$ was estimated both via an analytic error propagation procedure adding the separate contributions in quadrature and via a numerical Montecarlo analysis. Both analyses resulted in a systematic error on $T_{eff}$ of 18K for the equilibrium data (mainly due to the errors in the measurements of the capacitances at the amplifier input and in the oscillator mass) and of 20-30K for the nonequilibrium data in the different states. The largest error contribution in the nonequilibrium $T_{eff}$ is due to the calibration error in $\Delta T$, the error in the thermal expansion coefficient $\alpha$ and in the oscillator mass $m_l$. The effective temperatures are also affected by a random error (of size between 4K and 7K) originating from the error in the estimates of the area and resonant frequency of the curve fitting the resonance peak.

The systematic errors in the effective temperature ratio shown in Fig.5 and listed in Table S2 were also computed analytically and numerically, by considering the full expression of the ratio of effective temperature in a nonequilibrium state over the effective temperature in an equilibrium state. We note that in the temperature ratio a few quantities cancel out: this is the case for instance of the amplifier gain $G$ and of the oscillator mass $m_l$. Thus the temperature ratio is independent of these quantities and hence of their errors.

FEM study of the sensitivity for transverse vibrations

We did not attempt to calibrate our signal in terms of vibration of the mass at the bottom end of the rod. Thus in Figure 5 we assumed that the oscillator equivalent to the transverse mode was dominated by thermal noise. However we did investigate how the signal changes when the gap shrinks in response to an applied thermal difference. We set up an electrostatic model of the capacitor formed by the mass at the rod bottom and the fixed electrode: we did include

in the model the rounding of the mass edges that we realized so to diminish edge effects. In the model the mechanical parts (mass and electrode) were centered in a sphere of 0.2m radius, with unitary relative permittivity.

We varied the gap $g$, i.e. of the distance between the facing surfaces, in the range (50÷300)μm in steps of 10μm. Then we changed the mechanical model by tilting the fixed electrode along the axis laying on its top surface (i.e. surface facing the mass) and passing through the middle point of its longer edge. We quantify the tilt by the parameter $t$ which represents the change in height of the electrode at the edges of the mass along a perpendicular to the bottom surface of the mass: thus the tilt angle is the ratio of the tilt $t$ over the length $w$ of the edge of the mass bottom surface (47.5mm). For each gap value $g$ we changed $t$ in steps of 2.7μm starting from 0 μm up to the value 0.54×$g$. For each set ($g$, $t$) we meshed the model excluding the volume inside the mass and the electrode. The mesh was formed by about $6.5 \times 10^4$ tetrahedrons, of variable size: mesh elements were denser close to surfaces of electrodes and mass. Hence for each set ($g$, $t$) we computed the capacitance C($g$, $t$) via FEM. Figure S5 summarizes the results of the FEM computation: the ratios of the capacitance at a given ($g_i$, $t_j$) over the value corresponding to the parallel plate capacitor with the same gap $g_i$ are plotted against the height change in correspondence of the mass edges. The plot shows only the data in the range of interest to the results shown in this paper, i.e. $g$=(50÷150)μm: at the maximum applied thermal difference the gap shrinks to 62μm from an equilibrium value of 90μm. In the parameter range under study the capacitance of the tilted sensor changes by up to 2% with respect to the parallel plate case.

In alternative to the numerical method, we also estimated the capacitance of the non-parallel plate capacitor via analytic method. In fact one can write:

$$\frac{C(g,t)}{C(g,t=0)} = \frac{g}{w}\int_{-w/2}^{+w/2}\frac{dx}{g-\frac{tx}{w}} \approx 1+\frac{t^2}{12g^2}+\frac{t^4}{80g^4}$$

To compare the FEM results with the above, we fitted the data shown in Fig. S5 with a 2$^{nd}$ order polynomial $a_0+a_1 x+a_2 x^2$ obtaining $a_0=1$, $a_1=0$ and $a_2=0.079$ in good agreement with the analytic expression.

In the case of the transverse mode, we expect the tilt to assume much smaller values if the thermal noise is the dominant driving force. Thus, we are confident that our results are correct if we add a 2% systematic error in the sensitivity of the readout in the nonequilibrium states.

**Supplementary references**

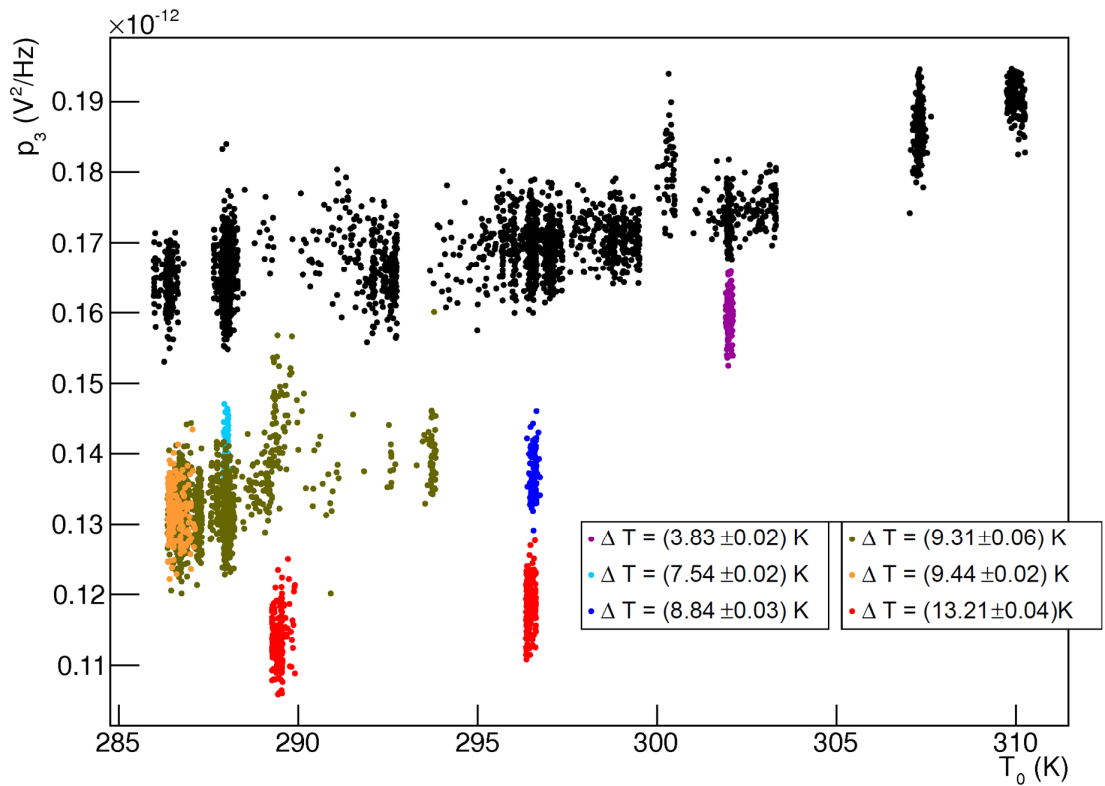

**Fig. S1** Plot of the constant $p_3$ used in the fit of the average PSD around the longitudinal mode of payload C as function of the temperature $T_0$ of the mass where the amplifier is fixed to. The different colours show measurements taken at different heat flows, hence different thermal differences: the colour code is shown in the legend.

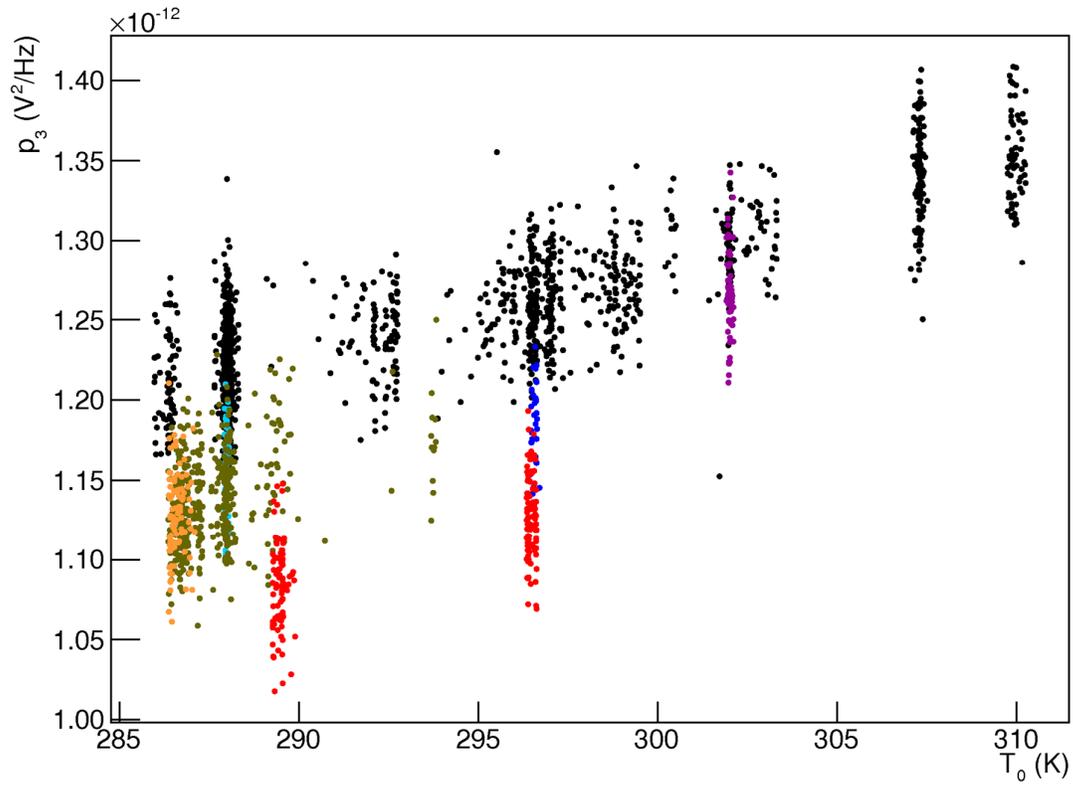

**Fig. S2** Plot of the constant $p_3$ used in the fit of the average PSD around the transverse mode of payload C as function of the temperature $T_0$ of the mass where the amplifier is fixed to. The different colours show measurements taken at different heat flows, hence different thermal differences. The colour code is the same as in Figure S1.

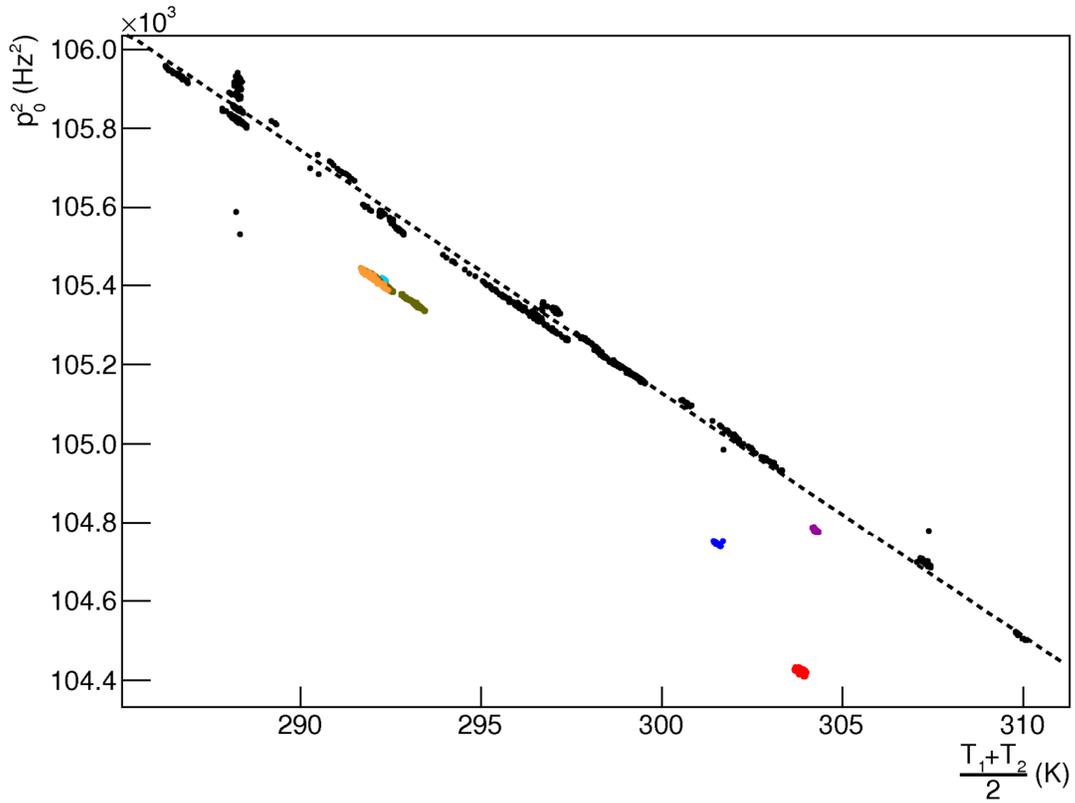

**Fig. S3** (Square of ) the resonant frequency of the transverse mode schematized in Figure 1c versus the average temperature of the rod extremes $(T_1+T_2)/2$. The black points refer to measurements in equilibrium states, while coloured points refer to NESS due to thermal difference $T_2$-$T_1$: the colour code is the same as in Fig. S1. The dotted line is the linear fit of the equilibrium data which we used to compute the average temperature of the mode.

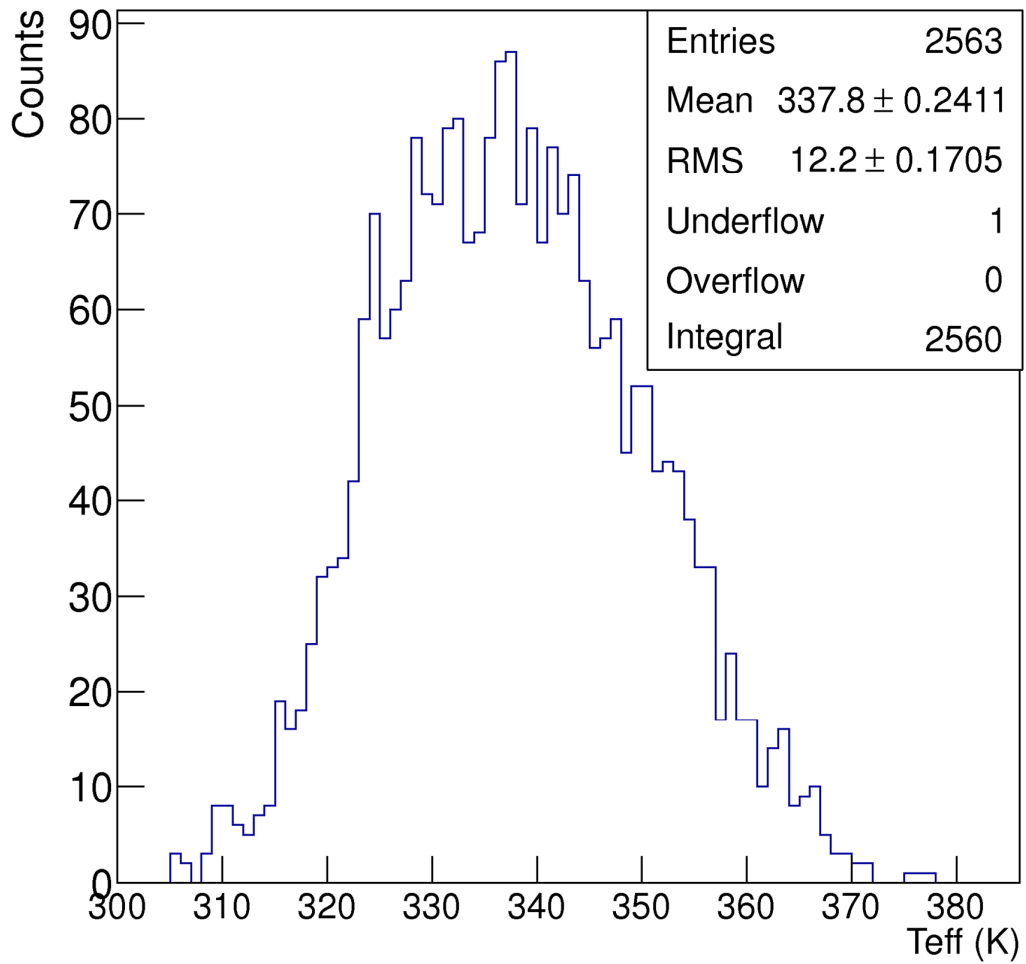

**Fig. S4** Histogram of the effective temperatures calculated in equilibrium for the longitudinal mode. The panel shows the statistics of the histogram. From this we infer an average effective temperature at equilibrium $T_{EQ}$= [338 ± 12(stat.) ± 18(syst.)]K

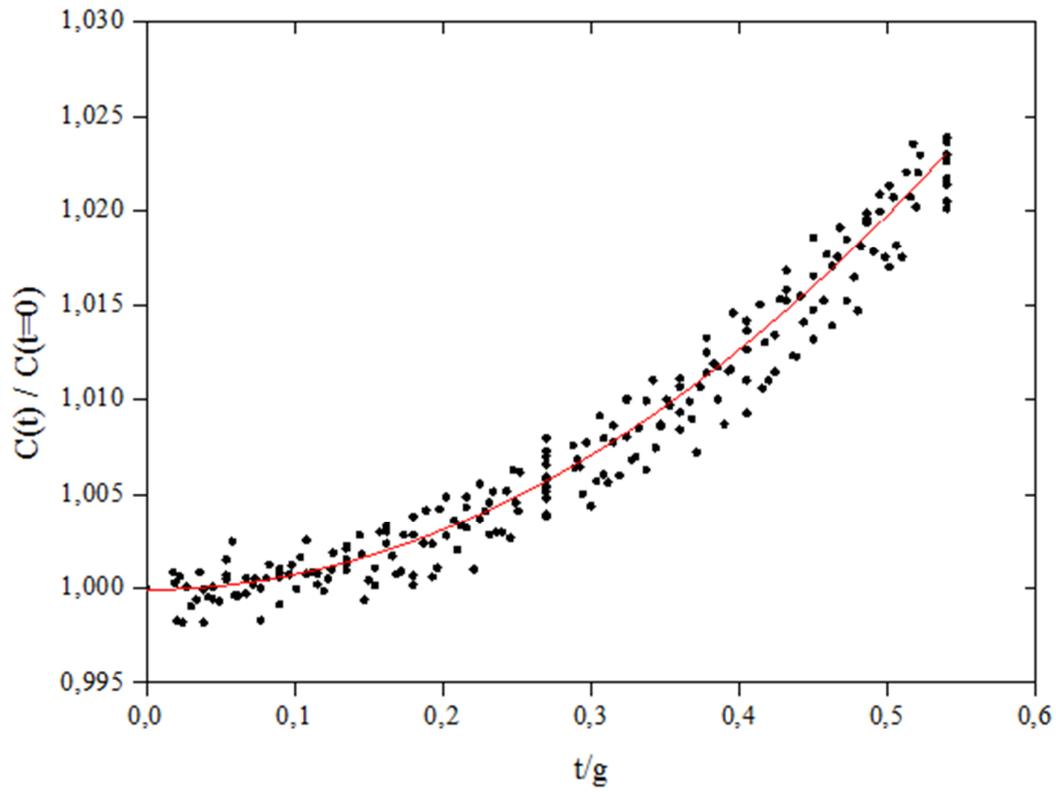

**Fig. S5** Change of the capacitance value between the mass and the fixed electrode, normalized to the value of the correspondent parallel plate case, as function of the electrode tilt $t$ at the mass edges divided by the gap $g$. The red line is the best fit with a $2^{nd}$ order polynomial.

| Sensor capacitance area | $(1599 \pm 2)\ 10^{-6}\ m^2$ | $C_{wire}+C_{testin}$ | $(35 \pm 2)\ 10^{-12}\ F$ |
|---|---|---|---|
| PTFE capacitance area | $(558 \pm 4)\ 10^{-6}\ m^2$ | Gain | $100 \pm 1$ |
| $C_0$ | $(284 \pm 2)\ 10^{-12}\ F$ | $m_l$ | $(0.234 \pm 0.008)\ kg$ |
| $C_{washer}$ | $(11 \pm 1)\ 10^{-12}\ F$ | $V_{bias}$ | $(value \pm 1)\ V$ |
| PTFE relative permittivity (Ref. S1) | $2.05 \pm 0.01$ | | |
| $\alpha$ (Ref. S2) | $(23.2 \pm 0.1)\ 10^{-6}\ K^{-1}$ | $\Delta T$ | $(value \pm 0.2)\ K$ |

**Table S1** Value and error of the independent quantities entering in the computation of $T_{eff}$; quantities on the last row affect only the effective temperature in the nonequilibrium states.

| ΔT (K) | R$_{EQ/NEQ}$ | |
|---|---|---|
| | Transverse mode | Longitudinal mode |
| 3.83 ±0.02(stat.) ±0.2(syst.) | 1.32±0.22(stat.) ± 0.03(syst.) | 1.04 ±0.036(stat.) ±0.013(syst.) |
| 7.54 ±0.02(stat.) ±0.2(syst.) | 1.76±0.29(stat.) ± 0.04(syst.) | 1.06 ±0.039(stat.) ±0.015(syst.) |
| 8.84 ±0.03(stat.) ±0.2(syst.) | 2.11±0.35(stat.) ± 0.05(syst.) | 1.13 ±0.041(stat.) ±0.017(syst.) |
| 9.31 ±0.06(stat.) ±0.2(syst.) | 2.14±0.35(stat.) ± 0.05(syst.) | 1.12 ±0.041(stat.) ±0.018(syst.) |
| 9.44 ±0.02(stat.) ±0.2(syst.) | 2.08±0.34(stat.) ± 0.04(syst.) | 1.11 ±0.040(stat.) ±0.018(syst.) |
| 13.21 ±0.04(stat.) ±0.2(syst.) | 3.31±0.55(stat.) ± 0.08(syst.) | 1.24 ±0.045(stat.) ±0.025(syst.) |

**Table S2.** List of the effective temperature ratio R$_{NEQ/EQ}$ between NESS and equilibrium states at different thermal differences for the transverse and longitudinal modes.



\end{document}